\documentclass[grl]{AGUTeX}

\usepackage{amsmath,graphicx,amssymb}

\setkeys{Gin}{draft=false}

\authorrunninghead{ROSAS-CARBAJAL ET AL.}

\titlerunninghead{3-D MUON TOMOGRAPHY OF LA SOUFRI\`{E}RE}

\authoraddr{Corresponding author: Marina Rosas-Carbajal, Institut de Physique du Globe de Paris. (rosas@ipgp.fr)}

\begin{document}


\title{Three-dimensional density structure of La Soufri\`ere de  Guadeloupe lava dome from simultaneous muon radiographies and gravity data}




\authors{Marina Rosas-Carbajal,\altaffilmark{1}
Kevin Jourde,\altaffilmark{2} 
Jacques Marteau, \altaffilmark{3}
S\'ebastien Deroussi, \altaffilmark{4}
Jean-Christophe Komorowski, \altaffilmark{1} and 
Dominique Gibert \altaffilmark{5}}

\altaffiltext{1}{ Institut de Physique du Globe de Paris, Sorbonne Paris Cit\'e, Univ. Paris Diderot, UMR 7154 CNRS, Paris, France.}
\altaffiltext{2}{BRGM, Orl\'eans, France.}
\altaffiltext{3}{Institut de Physique Nucl\'eaire de Lyon,
  Univ. Claude Bernard, UMR 5822 CNRS, Lyon, France.}
\altaffiltext{4}{Observatoire Volcanologique et Sismologique de
Guadeloupe, Institut de Physique du Globe de Paris, Paris, France.}
\altaffiltext{5}{OSUR - G\'eosciences Rennes (CNRS UMR 6118), Universit\'e Rennes 1, Rennes, France.}



\begin{abstract}

Muon imaging has recently emerged as a powerful method to
complement standard geophysical tools in the understanding of the
Earth's subsurface. 
Muon measurements can yield a ``radiography'' of the average density along
the muon path, allowing to image large volumes of a geological body from a
single observation point. 
Here we jointly invert muon data from three simultaneous telescope
acquisitions together with gravity  data to estimate
the three-dimensional density structure of the La Soufri\`ere de  Guadeloupe lava dome.
Our unique dataset allows us to achieve an unprecedented spatial
resolution with this novel technique. 
The retrieved density model 
reveals an extensive, low-density anomaly where the most active part
of the volcanic hydrothermal system is located, supporting previous
studies that indicate this region as the most likely to be involved in
a partial edifice collapse.

\end{abstract}



\begin{article}


\section{Introduction}

The La Soufri\`ere de  Guadeloupe volcano (Lesser Antilles) is an explosive subduction volcano holding one of the most hazardous volcanic hydrothermal systems in the world \citep{loughlin2015global}.  
It currently experiences an increasing unrest, raising concern on
possible pore-fluid overpressurisation and flank instability. 
Knowledge of the density distribution in its upper structure (lava dome) is crucial for hazard assessment since it can help to determine mechanically weak regions that may result in flank collapse.
Volcano density structures are traditionally studied with gravity
data, and 3D models solely based on these data are highly non-unique
and low-resolution, especially since field conditions often make it
difficult to achieve a good data coverage.

\begin{figure*}
\includegraphics[width=0.9\linewidth]{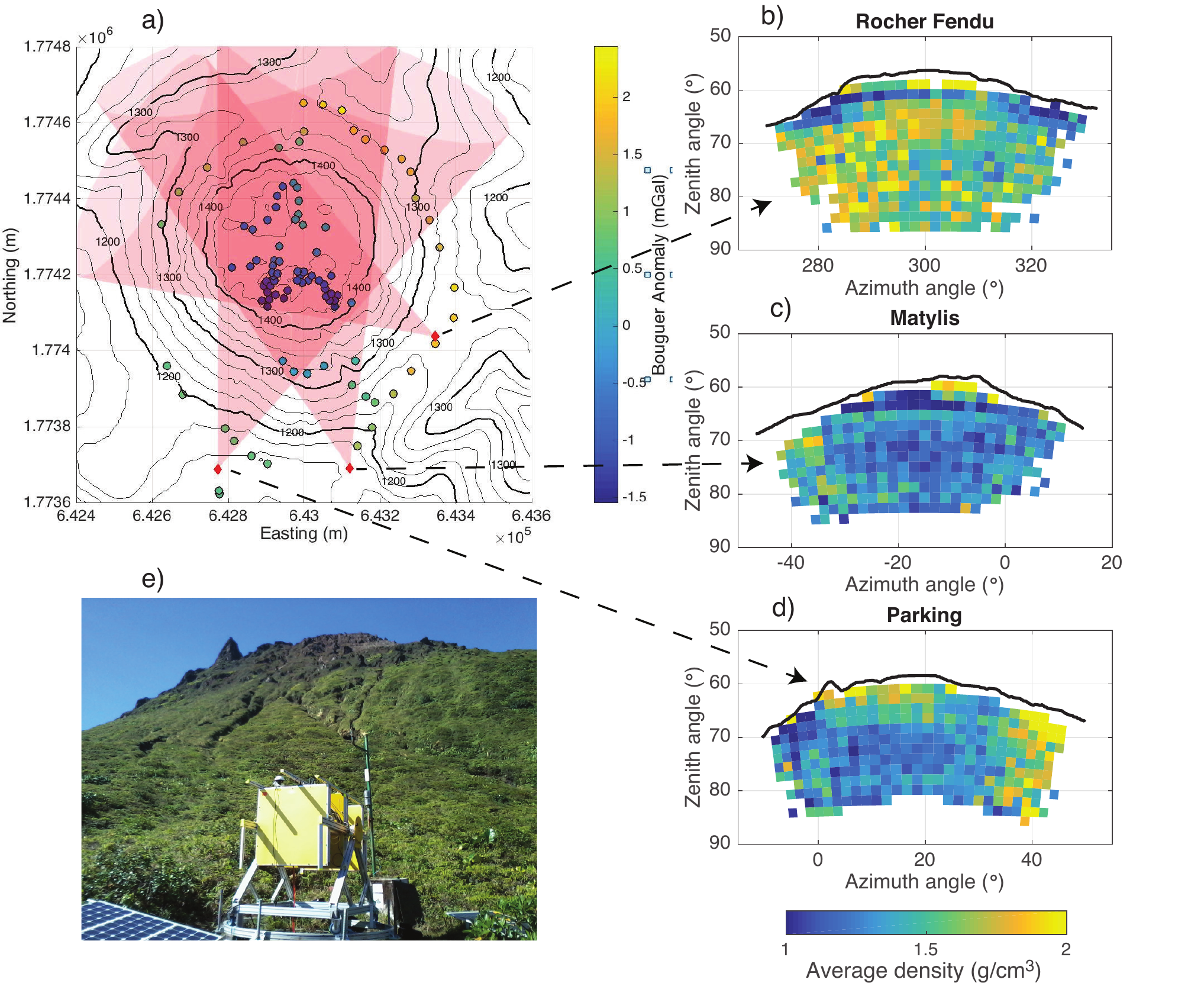}
\caption{ a) Location of muon telescopes (diamonds) and gravity stations (circles) used to reconstruct the La Soufri\`ere de Guadeloupe lava dome density distribution. Gravity stations are coloured according to their Bouguer anomaly. Red cones indicate the region of the dome sampled by each telescope. b-d) Muon radiographies obtained with the 3 telescopes. Each pixel corresponds to a particular line of sight of the telescope. The average density is computed from the opacity, which is inferred from the number of muons detected, and the total length of rock crossed by the muons. e) The muon telescope at the ``Parking'' site. }
\label{soufriereCoverage}
\end{figure*}

Since a decade, muon imaging has been developed by several teams to probe the density structure of various geological bodies \citep[e.g.][]{nagamine1995method, tanaka2001development,   lesparre2010geophysical}. 
The principles of the method are common to those of X-ray medical imaging, where particles are absorbed differentially according to the matter's density along their travel path \cite{nagamine1995method}. The method exploits the properties of muon interaction with ordinary matter, in particular a small cross-section and a quasi linear propagation path \citep{nagamine1995method}. Experimental muon trajectories are inferred from a series of pixelated particle detectors, such that the lava dome is sampled in hundredths of directions delimited by a cone (Fig. \ref{soufriereCoverage} a), and scanned typically in a few weeks \citep{lesparre2010geophysical}. The number of muons detected is used to compute the rock average density for each trajectory, resulting in 2D average-density radiographies (Fig. \ref{soufriereCoverage} b-d).

Among all projects using the muon imaging technique, DIAPHANE has been a pioneer since almost a decade in designing, producing and deploying portable, autonomous and robust detectors (so-called muon ``telescopes'') on the slopes of active volcanoes~: the La Soufri\`ere de Guadeloupe in the Lesser Antilles \citep{lesparre2012density,jourde2013experimental,jourde2016muon}, the Mayon in the Philippines, the Etna in Italy \citep{carbone2014experiment}. Despite difficult field conditions, data acquisition is continuous, deadtime-less with a high duty cycle (more than 95\%), and only limited by electrical power blackouts during the worst weather periods. Details on the methods and detectors developed by DIAPHANE may be found in \citet{lesparre2012design,jourde2013experimental, marteau_implementation_2014} and references therein. 

Information about the 3D density distribution of a geological body
may be obtained by combining more
than one telescope acquisition \citep[e.g.][]{tanaka2010three} or jointly inverting the muon data with gravity data  \citep[e.g.][]{nishiyama2014integrated,nishiyama20173d}.
\citet{nishiyama20173d} jointly inverted data from a single muon telescope with gravity data to study a volcanic lava dome. Their model has a limited spatial resolution due to the few data available (30 muon data points and 30 gravity data points). A study on the resolving kernels of muon and gravity data joint inversion by \citet{jourde_joined_inversion} suggests that in the case of an ideal coverage of the geological body by several muon telescopes, gravity data do not significantly improve the inversion results. However this might not be the case when the muon radiographies are obtained from a limited number of locations.

Here, we jointly invert muon data from three
acquisition sites together with gravity data points
(Fig. \ref{soufriereCoverage} a) to study the 3D density distribution
in the La Soufri\`ere lava dome. We first present the datasets, the
forward modelling and the inversion method developed to process the
muon and gravity data. We emphasize the sensitivity of the inversion
to the regularization scheme and its relation with the distance
between the measurement point and each voxel. 
In a second part we perform a synthetic test to cross-check the
methodology and constrain the regularization parameters. The last part
presents the results obtained in the 3D reconstruction of the dome and a discussion on the model's coherence with known geological structures and also identified limitations of the present data processing.


\section{Data acquisition and processing}

Muon data were simultaneously acquired by 3 muon telescopes at the ``Parking'', ``Matylis'' and ``Rocher Fendu'' sites (Fig.~\ref{soufriereCoverage} a) between June and August 2015, and for a duration of 63, 36 and 25 days, respectively. The measurement period was enough to obtain representative statistics of the outgoing muon flux in each direction. We applied the methodology proposed by \cite{jourde2013experimental} to filter out the upward flux of muons, which otherwise introduces a significant bias in the density estimation. The resulting average density as a function of the muon direction relative to the telescope orientation is presented in Fig. \ref{soufriereCoverage} b-d. 
The telescopes used in this analysis (Fig. \ref{soufriereCoverage} e) count $31 \times 31$ lines of sight, but in practice the number of data points per muon telescope is lower since some lines of sight point above the volcano or to regions where the rock thickness is too high. 
In our case, a total of 931 muon data points are obtained combining the 3 telescopes. We estimate the error for each muon data point as a combination of the measurement error and a model error related to the approximations done to solve the forward problem \citep{jourde2015thesis}.

The muon radiographies display density values that vary between 0.7
and 3 g.cm$^{-3}$, with average densities of 1.58 (Rocher Fendu), 1.32
(Matylis), and 1.43 (Parking)  g.cm$^{-3}$. These values are 
low compared to the typical density of 2.7 g.cm$^{-3}$ found
for andesitic rocks at La Soufri\`ere \citep{Komorowski2008}. Rock
alteration and dissolution caused by hydrothermal fluids, and the
presence of large fractures and cavities in the dome may strongly
reduce the bulk density of a lava dome as reported by
\citet{ball2015hydrothermal,voight200226}. 
Indeed, using a new analysis of data from the last (1976) phreatic
eruption \citep[c.f.][]{rosas2016volcano}
we re-estimated the total ejected volume of this eruption
to be 5-6 $\times10^6$ m$^3$. This large amount of
non-magmatic material implies that an extensive volume of the dome is
highly altered, and that large cavities prevail. 
\citet{gailler2013crustal} estimate the average density of the La
Soufri\`ere complex to be 1.8 g.cm$^{-3}$ based on regional gravity
surveys.
A low average density is thus not surprising,
however we also point out the existence of a 
diffusive flux of muons, which can bias the estimated density towards lower
values \citep{nishiyama2014integrated,nishiyama2016monte}. 
This phenomenon is currently being studied with dedicated simulations, which show that in the case of La Soufri\`ere it can effectively 
 shift the inferred densities to lower values. We therefore treat the
 densities displayed in the radiographies as relative rather than
 absolute. 
Low- and high-density regions can be clearly identified in the three
radiographies, even though no spatial smoothing has been applied to
the data processed.  

A total of 109 gravity measurements were done between March 2014 and
February 2015 with a Scintrex CG-5 gravimeter. Measurements have a
precision of the order of $10 \, \mu\mathrm{Gal}$. Details of the data
processing prior to inversion, including drift, tides, ellipsoid and
topography corrections are given by \cite{jourde2015thesis}. After
processing, we retained 103 good-quality data points to perform the
inversion. The correlation of the Bouguer anomaly with respect to
topography evaluated for different trial densities
\citep{jourde2015thesis,nettleton1939determination} is minimum for an
optimal Earth density of $1.75 \, \mathrm{g.cm}^{-3}$
\citep{parasnis1952study}. This value is smaller than the estimate
given by \cite{gunawan2005gravimetrie} in their regional study
comprising data points from the whole island of Guadeloupe, but we 
use our local estimation to focus on the density distribution
restricted to the La Soufri\`ere lava dome. The resulting Bouguer
gravity anomalies are presented in Fig.~\ref{soufriereCoverage}
a. Lowest gravity anomalies are found in the southern region of the
summit whereas high gravity anomalies are localized at the north and
west sides of the dome's base.

\section{Forward modeling and Inversion}

The muon forward problem involves calculating a convolution of the
incident muon open-air flux, their energy loss along the trajectory
inside the target, and the telescope acceptance
\cite[e.g.][]{lesparre2012bayesian}. 
The relevant parameters are, on the target side: the opacity, $\varrho
= \int_{L} \rho(\xi) ~ \mathrm{d}\xi $ (where $\rho$ is the density
and $L$ the particle path length) and on the probe side: the incident
muon flux. The differential muon flux associated with the line
$\mathbf{t} = \left(\mathbf{r},\varphi,\theta\right)$ ($\mathbf{r}$ is
the measurement point and $\varphi$ and $\theta$ are the azimuth and
zenith angles of the trajectory, respectively) may be expressed as a
function $\delta\phi_\mathbf{t} = \frac {\partial^3 \phi}
{\partial\Omega \partial S} (\varrho, \varphi, \theta) \,
[\mathrm{s}^{-1}\mathrm{cm}^{-2}\mathrm{sr}^{-1}]$ ($\Omega$ is the
solid angle and $S$ the surface) that accounts for the muon flux
reaching the instrument. 
This muon flux can be computed analytically at the surface of the Earth \cite[e.g.][]{tang2006muon, shukla2016} but, here, we use a more precise model obtained with the Corsika simulation
code \citep{heck1998corsika} which accounts for the altitude of the
lava dome. To simulate the energy loss of the muons interacting with
matter for a given opacity, we use the Monte-Carlo transport code
Geant4 \citep{agostinelli2003geant4}. 
Finally, to estimate the outgoing flux from the number of muon hits measured we use an experimental
acceptance function, calibrated on the field for each telescope \citep{lesparre2012bayesian}. 

Numerical tests using Geant4 show that the relation between observed opacities and
outgoing flux for each telescope line of sight can be treated as linear in the +/- 2
standard deviations range. Our linear
inversion fits the data within the +/- 1 standard deviation range. 
We thus use the relation between observed outgoing flux and opacities
for +/- 2 standard deviations to estimate
the Jacobian around this point. We then need to compute the opacity
resulting from a 3D density distribution. For this we discretize the
lava dome in cubes of $8\times 8 \times 8 \, \mathrm{m}^3$ using a
digital elevation model with a $5 \, \mathrm{m}$ precision and compute
the relative contribution of each cube to the total opacity for each
muon trajectory.  

For the gravity case, we decouple the problem between the region close
to the measurements and the region outside, used to compute the
topography correction \cite[c.f.][]{jourde2015thesis}. 
In the vicinity of the volcano we use a polyhedral discretization based on the 5 m resolution digital elevation model. We compute each polyhedra contribution to each gravity station with the program by \cite{d2013evaluation,d2014analytical}.

\begin{figure*}[!ht]
\includegraphics[width=1\linewidth]{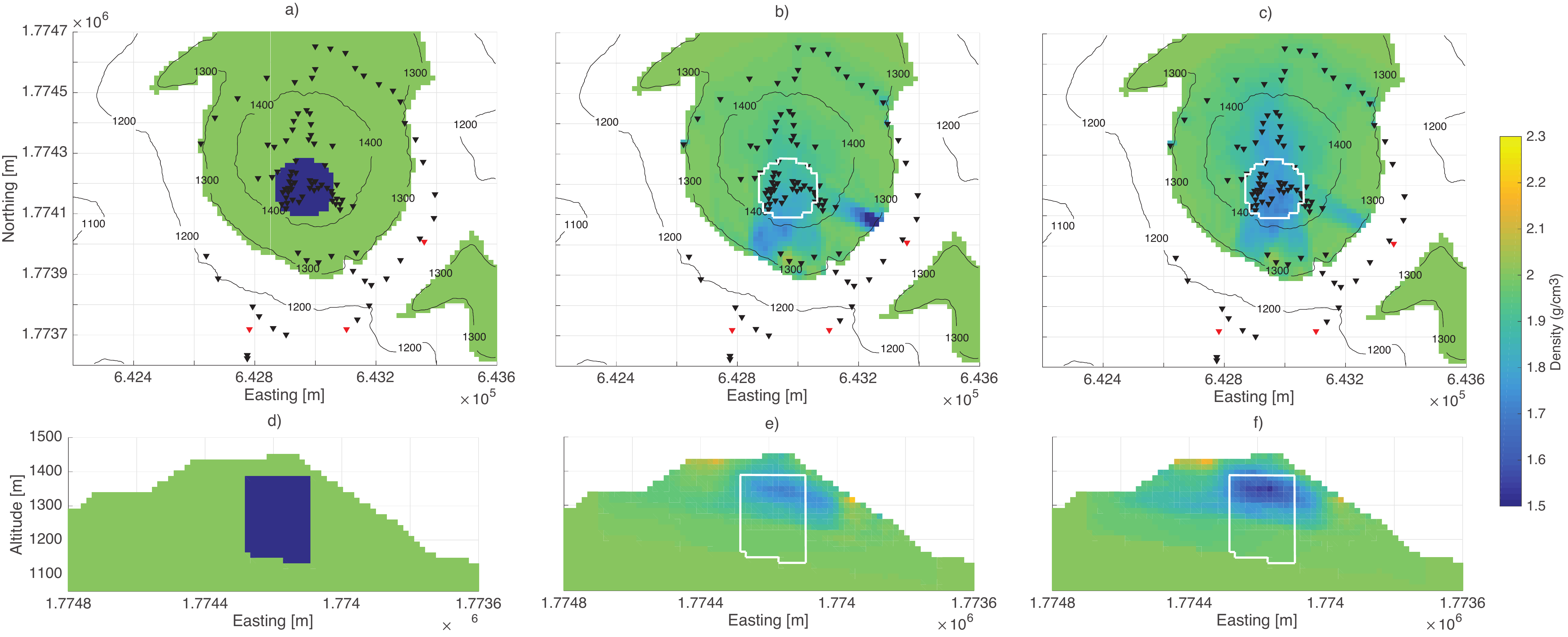}
\caption{Results of the synthetic data inversions used to study the
  scaling applied to the regularization matrix. Columns correspond to
  different slices of the density models. a,d) True model used to
  generate the synthetic data. b,e) Density model obtained from an
  inversion without applying any scaling to the regularization
  matrix. c,f) Density model obtained from an inversion using the
  regularization matrix scaling. Both models fit the data to the same
  level.}
\label{Fsyn}
\end{figure*}

The muon and gravity data are sensitive to the same physical parameter, thus a joint inversion can be
achieved by combining their forward kernels. We use the same model parametrization for the gravity and muon problems such that 
\begin{equation}\label{eq_lin2}
\text{G}  \begin{bmatrix} \boldsymbol{\rho}_\mu \\
\Delta \rho \end{bmatrix}
=  \begin{bmatrix} \text{G}_\text{g}
\\ \text{G}_\mu \end{bmatrix} \begin{bmatrix} \boldsymbol{\rho}_\mu \\
\Delta \rho \end{bmatrix} = 
\begin{bmatrix} \bf{d}_\text{g} \\ \bf{d}_\mu 
 \end{bmatrix} = \bf{d},
\end{equation}
where $\text{G}$ is the forward kernel, $\boldsymbol{d}$ the data vectors, subscripts $\text{g}$ and  $\mu$ denote the gravity and muon case, respectively, and $\boldsymbol{\rho}_\mu$ is the density
distribution discretized in cubes of $16 \times 16 \times 16
\mathrm{m}^3$ (forward kernels are built such that the same
discretization is used). The scalar parameter $\Delta \rho $ accounts for a possible density offset between the gravity- and muon-inferred models due to the negative density bias caused by the diffusion of low-energy muons on the volcano surface. Accounting for this offset, the density models are related through $\boldsymbol{\rho}_g = \boldsymbol{\rho}_\mu + \Delta \rho  $. Our linear inversion minimizes an objective function consisting of weighted data misfit term and a model regularization term:
\begin{equation}\label{eq_cost}
\begin{split}
\phi(\bf{m}) =
(  \bf{d} - \text{G} \bf{m})
^\text{T}\text{C}_\text{d}^{-1} (  \bf{d} - \text{G}
\bf{m}) + \\
+ \epsilon^2 (
\bf{m} - \bf{m}_\text{prior} )
^\text{T} \text{C}_\rho^{-1} (  \bf{m} -
\bf{m}_\text{prior} ),
\end{split}
\end{equation} 
where $\bf{m} = [ \boldsymbol{\rho}_\mu ,
\Delta \rho ]$, $\bf{m}_\text{prior}$ is the a priori density model
and density shift,
$\text{C}_\text{d}^{-1}$ is the data covariance matrix,
$\text{C}_\rho^{-1}$ is the model regularization matrix
and $\epsilon^2$ the trade-off parameter, which establishes the
relative weight of the regularization in the cost function
\citep{menke2012geophysical}. In practice, we adjust this parameter to
obtain the largest possible contribution of the model regularization
while fitting the data to the error level.
To determine the prior model we searched for the homogeneous density value
that minimizes the density offset. We found
that a value equal to the Earth density used to compute
the Bouguer anomalies results in a minimum offset. Accordingly, we
used a prior value of zero for the density offset.
The model regularization consists of a combination of smoothness constraints,
achieved by penalizing the difference in density between adjacent
cubes, and damping, which favours density models that are close to
$\bf{m}_\text{prior}$. 

Being sensitive to the opacity, muon data have no inherent information
on the density distribution along the muon path. Similarly, gravity
data have no inherent depth resolution. It is known that in the
regularized inversion of potential field data the lack of depth
information causes the retrieved structures to concentrate near the
surface. This problem can be overcome by using a scaling of the
regularization matrix to counteract the natural decay of the
sensitivity with depth. Since the original application on the
inversion of magnetic data by \cite{li19963}, this scaling has been
applied to other types of potential field data. Here, we combine a
depth scaling as introduced by \cite{li19963} to compensate for the
decay of sensitivity in the gravity data, with a distance-to-telescope
scaling to compensate for the same effect in the muon data. Details on
these scalings are given in the Supporting Information
\citep{boulanger2001constraints,ghalehnoee2017improving}. 
We emphasize
that this is the first time that such scalings are implemented in an
inversion of muon data. Previous studies deal with this problem by
reducing the amount of damping in the zone where anomalies are
expected, thus severely biasing the result of the inversion.


\section{Results and interpretation}

\subsection{Synthetic tests}

Before inverting the real data, synthetic tests are performed to
assess the influence of the regularization scaling and the resolution
that can be expected. Figure \ref{Fsyn}a and d) show slices of one of
the models used to generate the synthetic data contaminated by errors
with the same standard deviation as estimated for the real
data. Without scaling, the inversion places the largest density
contrasts close to the muon telescopes (Fig. \ref{Fsyn} b and e). In
turn, the matrix scaling places the density contrasts where the true
anomaly is located (Fig. \ref{Fsyn}c and f). Other tests with smaller
and larger radii of the cylinder and with its center close to one of
the muon telescopes did not change these results. Since the three
telescopes point northward, there is a more pronounced
under-determination in the North-South direction which results in a
spreading of the density anomaly in this direction. Inversions without
the gravity data (tests not shown) confirm the importance of these
data in the delimitation of the density anomalies in cases where the
spatial coverage of the telescopes is not complete.

The retrieved anomaly has a shorter depth extension than the true one
(Figs. \ref{Fsyn} d and f). Indeed, the position and inclination of
the telescopes determine a plane above which the density is
explored. In our case, the telescope in the Parking site is the one
with the lowest altitude. Its lowest plane can be inferred from
Figs. \ref{Fsyn} f where the anomaly begins. We therefore interpret
the density models only above this plane. Furthermore, given the large
non-uniqueness of the problem and the regularization used, we only
interpret our model in terms of low- and high-density regions without
asserting their precise delimitations. 

\subsection{Inversion of real data}

\begin{figure*}[!hb]
\includegraphics[width=1\linewidth]{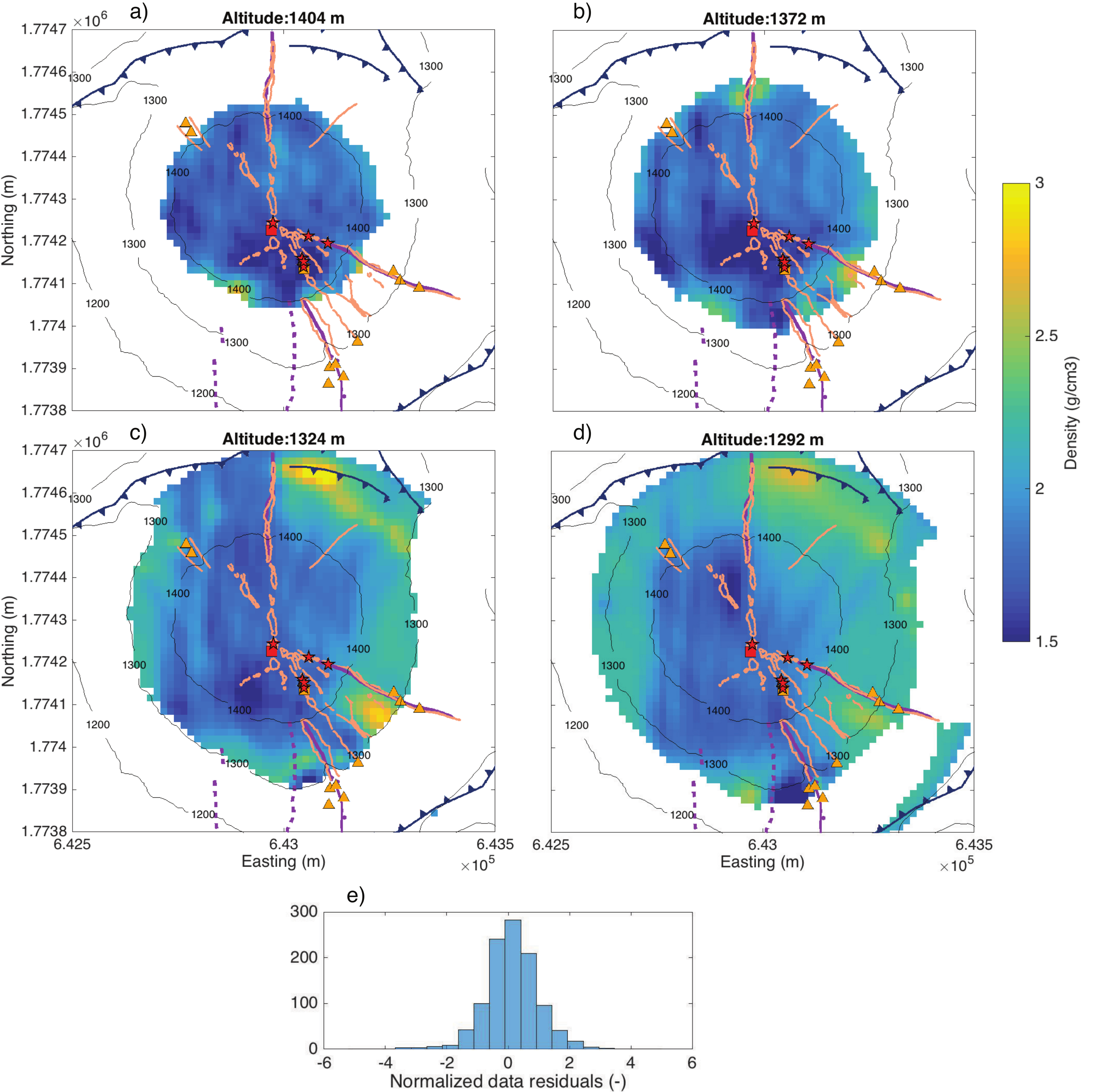}
\caption{a-d)  Horizontal slices of the 3-D density model obtained from
  the joint inversion of field muon and gravity data. A structural map
  is superposed to facilitate
  interpretation (orange lines: fractures; violet lines: faults; blue
  lines: collapse scars; orange symbols: past activity; red symbols:
  present activity; triangles: hydrothermal fluid exurgence;
  stars: active fumaroles; squares: boiling acid ponds). e) Histogram of the data residuals (gravity and muons) normalized by their error standard-deviation. Residuals add to a RMS
error of 1. }\label{Freal}
\end{figure*}

The density model resulting from the joint inversion of muon and
gravity real data is shown in Figs. \ref{Freal}a-d). The root mean
square error equals $1$ and the fit is of the same quality for both
the muon and gravity data (Fig. \ref{Freal}e). The offset parameter
equals $0.47 \, \mathrm{g.cm}^{-3}$. and the average density is of
$2.09 \, \mathrm{g.cm}^{-3}$, a more realistic value than $1.61 \,
\mathrm{g.cm}^{-3}$ retrieved from the muon data alone and consistent
with the average densities of Fig. \ref{soufriereCoverage}
b-d. Inverting the gravity and muon data sets without the density
offset parameter produces a model with density contrasts similar to
the model shown in Fig. \ref{Freal} down to an altitude of $\sim 1300
\, \mathrm{m}$. However, below this altitude and in order to improve the fit of the gravity data, the inversion places an artificial high-density region that mimics the role of the offset parameter.

In presence of strong topography variations, relative gravity data can
be used to retrieve an absolute density model \citep[e.g.][]{linde20143}. Simple synthetic tests
with anomalies like the one shown in Fig. \ref{Fsyn} suggest that the
altitude differences of $\sim 150 \, \mathrm{m}$ between the dome's
summit and base are sufficient to retrieve absolute density
values. Extending this conclusion to complex density
distribution is not straightforward and we prefer to restrain our
interpretations on relative density variations, although the densities
retrieved by the inversion fall in the expected range (see
Fig. \ref{Freal}).

The most extensive low-density region is found below the southern part of the
summit (Figs. \ref{Freal}a-c), where the most presently active
fumaroles are located. This region counts several deep (up to 122 m)
pits, craters and fractures that were active during the last 6
phreatic eruptions, as well as evidence of rock alteration
\citep{rosas2016volcano}. 
The low-density anomaly is well resolved since the three telescopes
are oriented to cover this part of the dome where many gravity
measurements were done (c.f. Fig. \ref{soufriereCoverage}a).
The low-density anomaly clearly extends down to $z = 1324 \,
\mathrm{m}$ but becomes unrecognisable at $z = 1292 \, \mathrm{m}$
level because of a limited resolution at this altitude level that is
only scanned by the telescope located at the Parking site. As a
consequence, the clear low-density anomaly observed for $z > 1324 \,
\mathrm{m}$ blurs into a faint low-density anomaly spread through the
imaged region (Fig. \ref{soufriereCoverage} d).

\begin{figure}[!ht]
\includegraphics[width=0.9\linewidth]{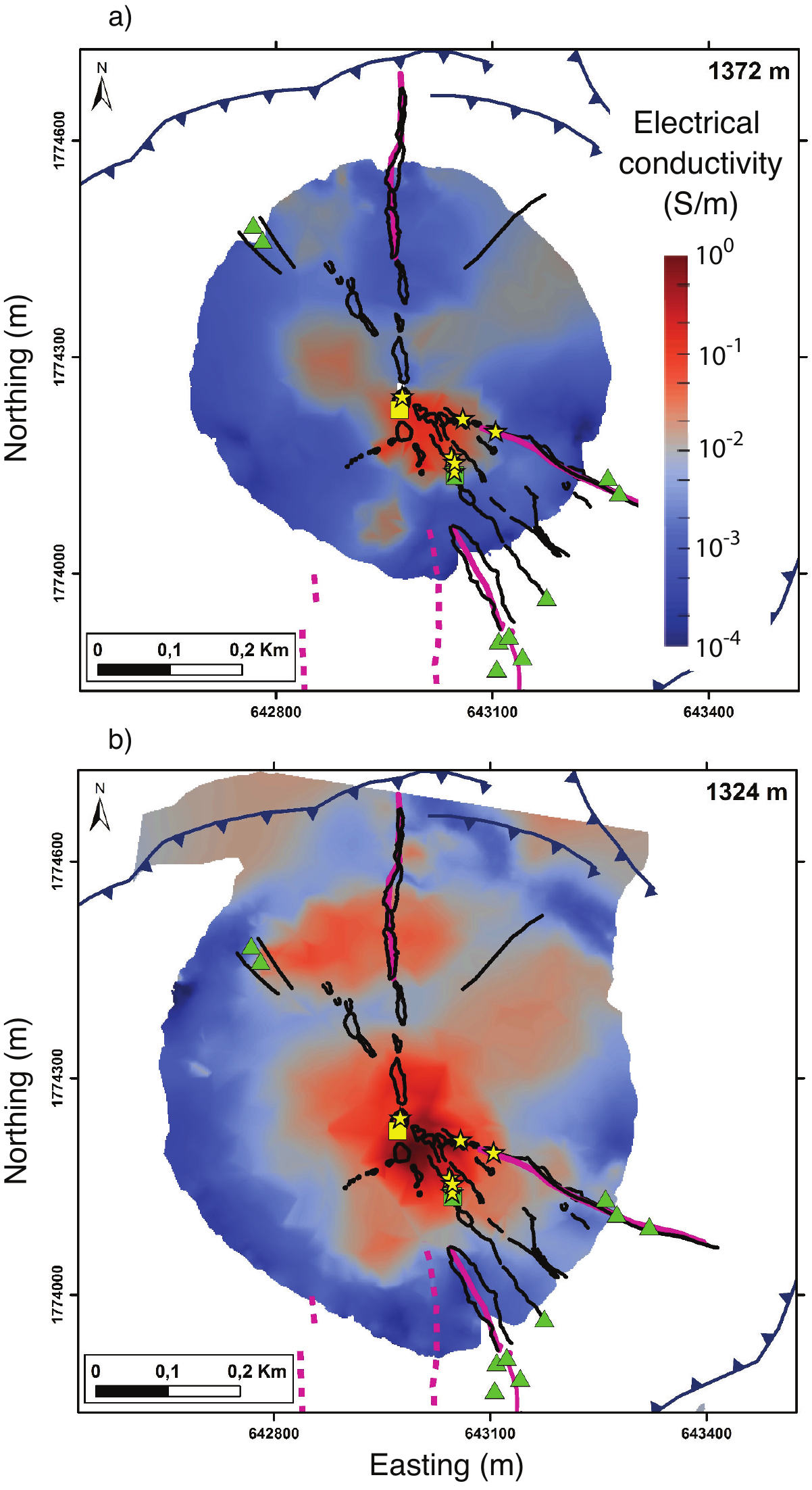}
\caption{Horizontal slices of the 3-D electrical  conductivity model by
  \cite{rosas2016volcano} at the same altitude as slices in
  Fig. \ref{Freal}b and c). High conductivity values correspond to hydrothermally
  altered, hot fluid-saturated rocks. Low conductivity values
  correspond to either cold, unaltered rocks, or regions where large
  cavities are present. See Fig. \ref{Freal} for map references.}\label{Fcond}
\end{figure}

The upper, well-resolved part of this anomaly coincides with the
electrically conductive region found by \cite{rosas2016volcano} with
the inversion of Electrical Resistivity Tomography (ERT) data. Slices
of the resulting electrical conductivity model at the same depths as
in Figs. \ref{Freal}b and d) are presented in Fig. \ref{Fcond}. While
the high conductivity values in this region may have been explained
by the presence of highly-conductive fluids (high temperature and
salinity), the low density confirms that this is also a
highly-altered, porous and thus mechanically weak zone, reinforcing
the likelihood of a partial edifice collapse hazard
\citep{rosas2016volcano}. Interestingly, the Tarissan boiling acid
pond (red square in figures) where the largest amount of magmatic fluids arrive to the surface is located in the northern limit of this low-density, high-conductivity region.

More limited low-density regions are found in the southwest-west close to
the summit (Figs. \ref{Freal} a-b, and in the extreme south and
north-center region at $\sim 1300 \, \mathrm{m}$
(Fig. \ref{Freal}d). The southernmost region is constrained to the
North by gravity data (c.f. Fig. \ref{soufriereCoverage}a) and it
probably corresponds to open fractures in the prolongation of La Ty
fault. 
This is supported by low conductivity values in the ERT model and 
historical evidence for exurgence of hot acid hydrohermal fluid from
this location (see orange triangle in Fig. \ref{Freal} d) during the
1976 phreatic eruption.
The low-density anomaly sub-parallel to the large north-south fracture
is located below a smaller, north-west fracture. Even if located relatively far for the muon telescopes, the
structure is well resolved thanks to the gravity points acquired in
its vicinity.
\cite{Kustersoufcavities} report on a large cavity located exactly in this
region.

High-density anomalies are found at the north-eastern base of the
dome (Figs. \ref{Freal}c-d), towards the southeast flank
(Fig. \ref{Freal}b-d), and in the south flank, west of the low-density
region (Fig. \ref{Freal}a-d). In the north, the anomaly correlates to
the position of the gravity measurements and follow the footpath along
the base of the lava dome. This anomaly may be structurally
controlled as it follows the detachment plane of an ancient collapse
\citep{rosas2016volcano}. Given the lack of data outside this region we do not
make further interpretations. In the south, southwest
(Fig. \ref{Freal}c-d), and in the northern extreme at $z = 1372 \,
\mathrm{m}$ (Fig. \ref{Freal}b) the high density coincides with low
electrical conductivity anomalies (Fig. \ref{Fcond}a-b), revealing an extensive
 volume of unaltered dome rock. The southern part of this anomaly
belongs to a ``bulge'', which rests on top of a highly altered zone
that reaches the surface in the South and thus constitutes a
potential detachment plane \citep[c.f.][]{rosas2016volcano}. 
Likewise, the high-density region
in the southeast flank corresponds to non-altered lava
dome spines. 

\section{Conclusions}

We obtained the first 3D density model of a volcano lava dome
from the joint inversion of simultaneous muon radiographies
and gravity data. Several density anomalies are detected in La
Soufri\`ere, especially an extensive low-density zone located below the
southern part of the summit where most of the increasingly active fumaroles are
observed. These results reinforce the inference from previous studies that the southern flank of
the La Soufri\`ere lava dome is highly porous and mechanically weak,
as well as saturated with hot acid fluid. 
The associated rock alteration and
dissolution promotes instability and raises the likelihood of a
partial collapse. 
Our 
density model constitutes the basis for undertaking monitoring of density
changes using simultaneous muon measurements from different points
around the lava dome. Future work will focus on improving the spatial
coverage to refine the model resolution. New recently installed 
muon telescopes on the west and northern parts of the
volcano are fitted with a smaller pixel support (63 $\times$ 63 lines of
sight). 
Theoretical work on the behaviour of the diffusive muon flux will be
crucial to estimate absolute density and its uncertainty. Joint inversion
with other types of geophysical data such as ERT data will be also
explored. 
This work demonstrates the potential of muon data to move from
2D ``radiographies'' to dynamic 3D imaging of geological structures. 
This is of paramount importance given the
limitation of conventional volcano monitoring in tracking
 the internal state of the
system with full spatial coverage.





%
%
%


\begin{acknowledgments}
This research was partly financed by the Swiss National Science Foundation
and is part of ANR DIAPHANE project ANR-14-ce04-0001. 
We acknowledge the financial support from the UnivEarthS Labex
program of Sorbonne Paris Cite (ANR-10-labx-0023 and
ANR-11-idex-0005-02) and from the Service National d'Observation en
Volcanologie (INSU).
We thank the team of the Observatoire Volcanologique et Sismologique
de Guadeloupe (OVSG-IPGP) for assistance with the gravity surveys. 
The installation of the muon telescopes benefited of the help of the crews
of the French Civil Security helicopter basis
(http://www.helicodragon.com). 
The authors declare no competing financial interests. 
The data used are displayed in Figure 1 a-d of this paper.
The CG-5 gravimeters were provided by the CNRS-INSU department. The
25 m resolution Digital elevation Model (BDAlti), building and terrain
information (BDTopo) were obtained courtesy of IGN (2004) whereas the
5 m resolution DEM was produced by IPGP/Latitude Geosystems using
stereo satellite imagery from GeoEye acquired on 20 November
2005. We thank Nolwenn Lesparre and the Editor Andrew Newman for their
guidance. This is IPGP contribution 3860.
\end{acknowledgments}











\bibliographystyle{agu}





\end{article}


%
%
%
%
%
%





\end{document}